# Graphene and thin graphite films for ultrafast optical Kerr gating at 1 GHz repetition rate under focused illumination

*Amr Farrag*[*,1], *Assegid M. Flatae*[1], *Mario Agio*[1,2]

1. Laboratory of Nano-Optics, University of Siegen, 57072 Siegen, Germany
2. Istituto Nazionale di Ottica (INO), Consiglio Nazionale delle Ricerche (CNR), 50019 Sesto Fiorentino (FI), Italy

*E-mail: amr.farrag@uni-siegen.de, mario.agio@uni-siegen.de

Keywords: ultrafast phenomena, optical non-linearity, third-order nonlinear process, graphene, graphite films, 2D materials

**Abstract**

The ability to address sub-picosecond events of weak optical signals is essential for progress in quantum science, nonlinear optics, and ultrafast spectroscopy. While up-conversion and optical Kerr gating (OKG) offer femtosecond resolution, they are generally limited to ensemble measurements, making ultrafast detection in nano-optics challenging. OKG, with its broadband response and high throughput without phase-matching, is especially promising when used at high repetition rates under focused illumination.

Here, we demonstrate an ultrafast detection scheme using the third-order nonlinearity of graphene and thin graphite films, operating at 1 GHz with sub-nanojoule pulses and achieving 141 fs temporal resolution. Their exceptionally large nonlinear refractive index, orders of magnitude higher than conventional Kerr media, enhances detection efficiency at smaller thicknesses, enables sub-picosecond response, and supports broadband operation. Their atomic-scale thickness minimizes dispersion and simplifies integration with microscopy platforms, optical fibers, and nanophotonic circuits, making them a compact, practical material platform for nano-optical and on-chip ultrafast Kerr gating.

## 1. Introduction

Understanding processes that occur on extremely short timescales is central to advancing ultrafast optical science, with implications ranging from quantum technologies to biological imaging and high-precision sensing [1–5]. Yet many excited-state transitions evolve so rapidly that they lie beyond the reach of traditional measurement techniques [6–7]. Moreover, advances in hybrid quantum systems, where quantum emitters are coupled to nanophotonic structures, have made sub-picosecond emission dynamics increasingly accessible experimentally [8–11]. These developments demand ultrafast techniques capable of probing photodynamics at the single-emitter level. While transient absorption, up-conversion, and optical gating approaches have been successfully applied to molecular ensembles [12–16], they are typically implemented at kHz or MHz repetition rates and are not well-suited for individual quantum emitters. In contrast, quantum-optical platforms, such as hybrid single-photon sources with sub-picosecond emission, require high repetition rates to achieve meaningful signal-to-noise ratios and accurately characterize photon statistics.

Ultrafast nonlinear optical sampling techniques based on fluorescence up-conversion, arising from second-order nonlinear interactions such as sum-frequency generation in nonlinear crystals, offers excellent temporal resolution across the ultraviolet [17], visible [18], and near-infrared [19] ranges. This technique has enabled studies of solvation dynamics [20], intramolecular vibrations [21], biophysical processes [22], and quantum applications [15]. However, its dependence on phase matching, narrow spectral bandwidth, and sensitivity to crystal parameters and group-velocity mismatch limits its efficiency for ultrafast single-photon detection, although single-photon frequency conversion has been demonstrated [4].

In contrast, optical Kerr gating (OKG) achieves femtosecond resolution without phase-matching constraints and operates over a broad spectral range. Time resolutions below 100 fs have been reported using bismuth glass [23] and fused silica [12]. Moreover, Kerr efficiencies exceeding 90% (up to 99.7%) have been demonstrated in YAG, GGG, BGO, and fused silica [24], making OKG a promising approach for high-efficiency, ultrafast single-photon detection.

Traditional Kerr media that are mentioned above provide femtosecond temporal resolution but suffer from low third-order nonlinearities, requiring high pump energies or long interaction lengths to achieve efficient gating. These constraints limit operation speed, scalability, and integration with compact optical systems. Graphene and graphite thin films overcome these limitations by offering third-order nonlinear refractive indices orders of magnitude higher than those of conventional media, enabling efficient Kerr gating even with sub-nanojoule pulses at high repetition rates. Their ultrafast carrier dynamics support sub-ps response times, while their broadband optical response allows operation across a wide spectral range without phase-matching constraints. Being atomic-scale thin, these materials introduce minimal dispersion and can be easily integrated onto optical fibers, waveguides, or nanophotonic chips. Consequently, graphene/graphite film-based Kerr gates provide an efficient, broadband, and scalable platform for ultrafast single-photon detection and time-resolved optical measurements.

In this manuscript, we demonstrate an ultrafast OKG operating at a 1 GHz repetition rate with sub-nanojoule pulse energies by focusing gate and probe on graphene and thin graphite films, following the proposal of Ref. [25]. Below we present the sample preparation and characterization, experimental approaches, and discussion.

## 2. Sample fabrication and characterization

Several graphene and graphite films were fabricated on glass substrates using a modified mechanical exfoliation technique, in which layers are peeled from a bulk graphite crystal. For their use as optical Kerr media, it is essential to obtain large-area flakes, on the order of several tens of micrometers, with minimal defects. Small flakes tend to accumulate heat during operation, as their limited area restricts thermal dissipation, increasing the risk of laser-induced damage. Larger flakes, in contrast, dissipate heat more effectively and are therefore better suited for optical Kerr gate experiments. To achieve large and high-quality flakes, we adopted the approach of Huang et al. [26], incorporating modifications to suit our equipment and to allow direct exfoliation onto glass substrates. The preparation began by cleaning the glass slides: first by rinsing them in distilled water and isopropanol for 10 min in an ultrasonic bath, followed by a 10 min UV-ozone treatment on each side to remove residual organic contaminants. These cleaning steps can be extended if needed. Exfoliation should begin immediately after cleaning to minimize surface contamination.

For exfoliation, we used 19 mm × 33 m Scotch Magic Tape (3M). The choice of tape and substrate is crucial for obtaining large-area flakes. A long strip of tape was first pressed onto the graphite mesa to lift a thin graphite sheet. The sheet was then folded and peeled several times to reduce its thickness before pressing the tape onto the glass substrate, where it remained during the fabrication process. To avoid contaminating the final flakes, we first removed the outermost graphite layers using a small piece of tape before carrying out the main exfoliation. Based on the fragmentation behavior discussed in Ref. [27] and the guidelines of Ref. [26], peeling the graphite on the tape approximately four times provides an optimal balance, producing large-area graphene flakes while avoiding excessive fragmentation into small, unusable pieces.

After transfer, the samples were heated to ~100 °C for 2 min and allowed to cool down to room temperature. The tape was then removed with a rapid peel, leaving behind large graphene and thin graphite films on the glass substrate. Examples of the fabricated samples are shown in Fig. 1a and b.

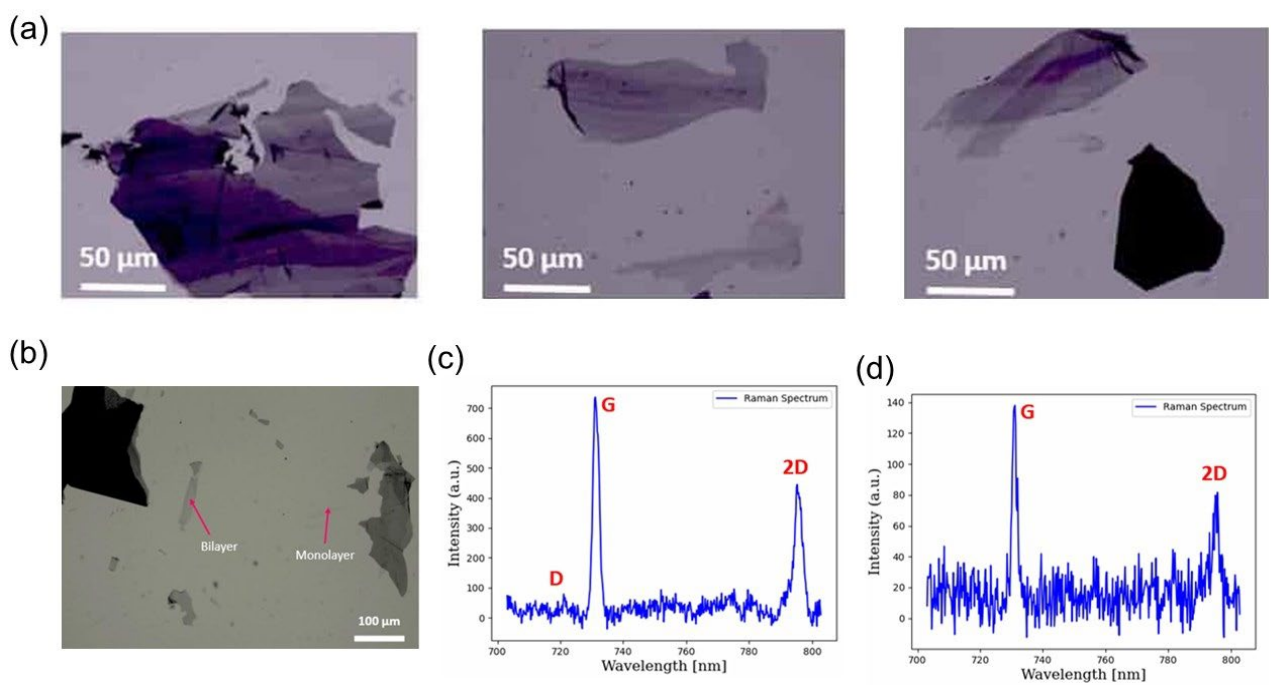

**Figure 1.** (a) Optical micrographs of graphite residues, thin graphite films, and graphene flakes on a glass substrate. (b) Identification of monolayer and bilayer graphene using optical transmission contrast. (c–d) Raman spectra of (c) a graphite flake and (d) a graphene flake, highlighting the characteristic G and 2D bands.

As shown in Fig. 1a, this approach yields large-area flakes. The dark black regions correspond to graphite residues and thicker graphite strips, while the lighter, grey-toned areas represent thin graphite films. Very thin, nearly transparent regions correspond to few-layer graphene. To determine the number of graphene layers, we employed a model adapted from Gaskell et al. [28], modifying it for optical transmission rather than reflection. Under an optical microscope, this method allowed clear visual distinction between different layer counts. Figure 1b, for example, displays a monolayer and bilayer graphene flake. It is worth noting that beyond approximately ten layers, the material no longer exhibits the characteristic properties of graphene and is better classified as a thin graphite film, which, as shown later, still produces a measurable optical Kerr signal.

To further characterize the samples, we performed Raman spectroscopy, a well-established and versatile technique for probing graphene's structural and electronic properties [29]. Raman measurements have been widely used to determine layer number [29], investigate electric-field–induced modifications to electron–phonon coupling [31], assess structural disorder [29, 32,33], and examine stacking order in multilayer graphene [34–37].

Our samples were characterized using a PicoQuant CW laser at 656 nm. Figure 1c and 1d show the Raman spectra of a graphite flake and a graphene flake, respectively. The characteristic G and 2D bands are clearly visible in both spectra. In the graphite sample, the disorder-induced D band is also present, indicating the presence of defects. In contrast, the graphene flake (10–11 layers) exhibits no detectable D band, suggesting a low-defect, high-quality structure, limited by the detection sensitivity of the used setup, as shown in Fig. 1d.

The exposure to highly intense light sources, such as fs lasers can cause damage to a bulk graphite, thin graphite films, or graphene. The damage can manifest itself in various forms, i.e., degradation of lattice like defects or cracks, or complete ablation of the carbon layers. Despite being detrimental, those deleterious effects can lead to controllable patterning of graphite or graphene, whenever the appropriate laser parameters are identified. This shows the importance of studying laser-induced damage of graphene or graphite films. For example, the controllable formation of nanopores in graphene crystal structure [38] could be an alternative way to engineer the electronic properties [39], chemical reactivity [40], and surface interactions [41]. Since graphene layers are isolated from bulk graphite (exfoliated graphene) or accumulated on top of each other (CVD-grown graphene), it is essential to understand the damage mechanisms associated with each physical system that are inter-related to one another. A theoretical analysis of the physical mechanisms of the damage formation induced by fs laser pulses in graphite films was performed by Jeschke et al. [42], using molecular dynamics calculations based upon a tight-binding Hamiltonian. This analysis identified a unique property of graphite, that is, it has two distinct laser-induced structural instability mechanisms, one at high laser energy and another at low laser energy. At high energy, e.g. $E_0 \geq 3.3 \pm 0.3$ eV/atom and a flux $F \approx 0.29 \pm 0.03$ J/cm$^2$, the fs laser excites electrons from occupied states to unoccupied states. Later, these electrons thermalize to a hot Fermi-Dirac distribution within ~10 fs, due to electron-electron scattering. This causes a rapid change of the interatomic potential leading to weakening of the in-plane C-C bonds. Within this very short time, the sp$^2$ in-plane C-C bonds will be broken. At 80 fs from the pulse maximum (pulse width = 20 fs), the disorder inside the in-plane graphite layers evolves with such high absorbed energy, and large volume expansion is observed. Some graphite planes will form strong covalent bonds (cross-linking), rather than being typically attached via the weak van der Waals forces, while some carbon monomers will be evaporated. This mechanism gives rise to melting and evaporation of graphite layers (graphene) from graphite films.

At low energy, e.g. $E_0 \geq 2.0 \pm 0.4$ eV/atom and a flux $F \approx 0.17 \pm 0.04$ J/cm2, the laser pulse excites strongly out-of-plane (c-axis) vibrations in graphene sheets. Typically, the planes of carbon atoms are separated by ~ 3.4 Å. But, due to those vibrations, the graphene layers come close by ~ 1.8 - 2.0 Å, whereas at such close distance the layers interact strongly with each other, with energy $E \approx 1$ eV, which results in collision and momentum transfer among them. In that situation, the surface layer with initially zero momentum acquires enough kinetic energy sufficient to escape the material. We see here that, with this mechanism, the graphene layers of the graphite film remain almost intact, with of course some disorder among layers, and removal of some surface layers. It was also found that the damage mechanisms are irrelevant to the laser pulse duration.

The damage caused by fs laser irradiation of graphene is commonly characterized by three main techniques: optical microscopy, Raman spectroscopy and atomic force microscope (AFM). In the work of Beltaos et al. [43], the damage effect of a Ti:Si laser at 840 nm, ~ 150 fs, with 76 MHz on

single- and multi-layer graphene samples, prepared via mechanical exfoliation was characterized using the three techniques. A laser interaction at ~ 350 mW power for several hours, led to a clear damage of graphite and a complete ablation of the single layer graphene, indicating graphite has a higher damage threshold. Investigation of another sample of 10 -15 layers, in the same work, showed degradation of graphene layers in some regions, while other regions suffered from an ablation with clear edges. Additional confirmation of the damage and degradation was obtained by taking the Raman spectrum before and after the laser irradiation, where the appearance of D peak is a clear indication of damage occurrence. Investigation of different laser power ranges 50 - 275 mW and exposure times (1 - 240 s), showed that the ablation threshold is ~ 4.2 mJ/cm$^2$ and the ablation power and exposure time can be as little as 250 mW and 2 s, respectively. Roberts et al. [44] studied the effect of very short laser (50 fs) to a single layer graphene fabricated by CVD technique and found out that it can withstand up to ~ 3 ×10$^{12}$ W/cm$^2$ (~ 200 × mJ/cm$^2$) from a single laser shot before it ablates. Also using 50 fs laser pulses and CVD grown single layer graphene, later transferred to a sapphire substrate, Currie et al. [45] conducted a quantitative study of laser-induced damage, and found that an optical fluence of 14 mJ/cm$^2$ leads to a local modification of sp$^2$ carbon bonding structures, with a linear increase in the damaged area when the laser fluence is increased. Further, at laser fluence of 57 mJ/cm$^2$ (180 mW average power, 1400 GW/cm$^2$ peak irradiance), the damage caused by laser irradiation, not only induced the D peak in the Raman spectrum but also caused disappearance of both G and 2D Raman peaks.

A time-dependent structural modification of single-crystal graphene layers, due to laser irradiation of as little as 1 mW power has been reported by Krauss et al. [46]. The exposure period to such modest power results in different modifications. For example, after 2 hours of 1 mW laser power, the heat removes the adsorbed dopants, like water and $O_2$, which manifests itself in the form of a redshift of G and 2D peaks, broadening of the G peak, and an increase in the $I_{2D}/I_G$ ratio. Exposure to similar power for hours to ten hours breaks the sp$^2$ C-C bonds, and results in disassembling the single-crystal into ~ 10 nm interconnected nanocrystallines. The Raman spectrum changes associated with that are strong D peak growth, a blueshift from phonon confinement, and a reduced 2D peak. Surprisingly, although the initial exposure to laser removes adsorbed dopants, the newly created grain boundaries act as docking sites for adsorbates. Additionally, p-type doping is observed for the laser exposed regions. Samples of bilayer graphene samples have shown similar results. Increasing the number of graphene layers weakens substantially this effect. Interestingly, bulk highly oriented pyrolytic graphite (HOPG) does not exhibit similar behavior, even at higher laser power of 12 mW. As far as we see, based upon the different experiments discussed above, seemingly there is no common agreement on an energy fluence value that can be considered the damage threshold for graphene. This might be because the studied graphene samples are fabricated under various conditions, and/or different substrate materials influence graphene samples differently.

In our samples, we observed damage after exposure to ~ 100 fs laser, 820 nm, 1 GHz repetition rate at 1 Watt average power (Taccor Tune 10, Novanta). Given the average power and the repetition rate, the laser energy is 1 nJ at the Kerr shutter. The beam size was 20 µm (area = 3.14 ×10$^{-6}$ cm$^2$), leading an energy fluence of ≈ 0.31 mJ/cm$^2$. Note that this laser fluence causing damage is nearly more than one order of magnitude lower than the lowest reported laser damage fluence (4 mJ/cm$^2$) in Ref. [43]. Figure 2 shows a graphene sample containing 10-11 layers before laser exposure (Fig. 2a) and after laser exposure at different sites (Fig. 2b). We used optical microscopy and Raman spectroscopy as tools in the determination of the damage threshold for thin graphite film and 7-8 layers of graphene (not shown). We observed that most samples remained intact after one hour of exposure to a 1 W fs laser, but some later developed cracks like Fig. 2b. Using the fs laser at lower powers (e.g., 290 mW) did not reliably prevent damage: the same thin films could survive under these conditions on some days yet crack on others. The reasons for this inconsistent behavior remain unclear. Possible factors include humidity-driven changes in the damage threshold of 2D materials, which are highly sensitive to adsorbed molecules [47]. Another possibility is fs-laser-induced structural weakening: ultrashort pulses can generate micro-cracks or voids, with heat promoting crack propagation, as reported for glass [48, 49]. A similar mechanism may occur in graphene. Further studies are required to clarify the parameters that govern fs-laser-induced damage.

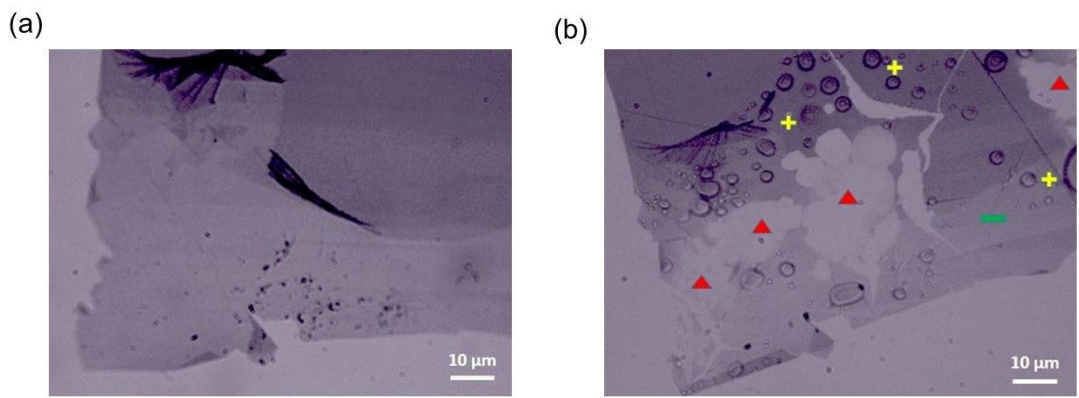

**Figure 2.** (a) Optical micrograph of a graphene flake with 10–11 layers. (b) The same sample after exposure to fs-laser irradiation. Red triangles mark regions of complete ablation, the yellow plus sign highlights crater-like damage spots, and the green minus sign indicates an area where the graphene has been partially thinned.

**3. Graphene and thin graphite film as a Kerr medium**

Graphene's linear energy-momentum dispersion gives it a strong nonlinear optical response over a wide spectral range, achievable with moderate electric fields [50]. This enables studies of harmonic generation, four-wave mixing, and saturable absorption [51, 52]. In addition, graphene's zero-gap structure enables tunable absorption over a wide spectral range, with low saturation intensity due to Pauli blocking. Graphene's third-order nonlinear susceptibility has been measured via four-wave mixing, showing large values (~$10^{-19}$ $m^2V^{-2}$) [53]. Its nonlinear response arises from both intraband and interband dynamics [54, 55].

For OKG experiments, it is essential to examine the self-phase modulation of graphene and thin graphite films to assess their ability to spectrally broaden the gate laser beam. To do this, several thin graphite films and graphene flakes were tested. For each sample, the gate beam spectrum was first measured without the sample, then the sample was inserted, and the spectrum was measured again for comparison. Figure 3 shows the spectra of ~ 8 layers of graphene (Fig. 3a), and thin graphite layer (Fig. 3b).

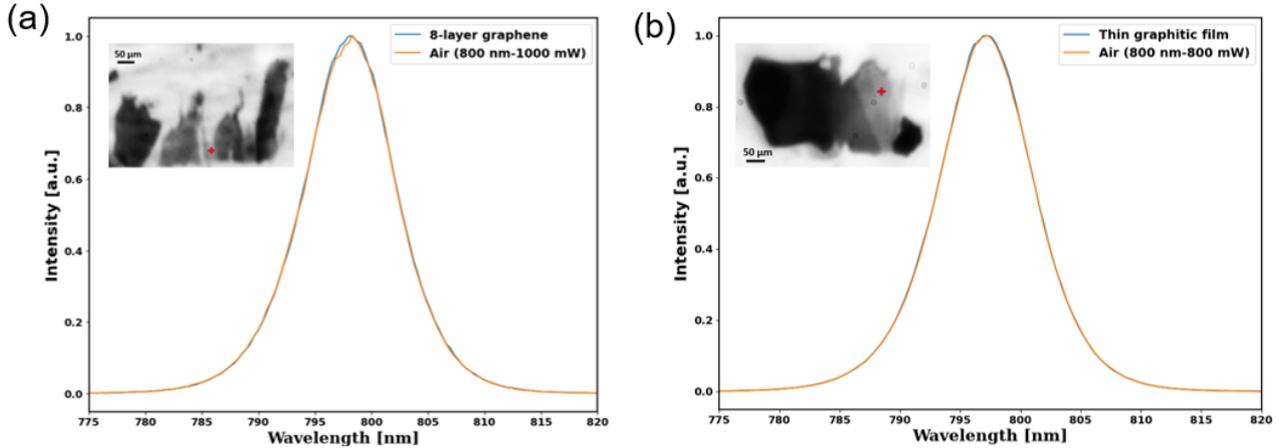

**Figure 3.** Self-phase modulation measurements for (a) an 8-layer graphene flake and (b) a thin graphite film. The laser wavelength is λ = 800 nm, with average powers of 1000 mW (pulse peak intensity ~ 1.33 GW/cm$^2$) for the graphene sample and 800 mW (pulse peak intensity ~1.06 GW/cm$^2$) for the thin graphite film.

The measured samples show no spectral broadening due to self-phase modulation. The broadening reported in Ref. [56] is not observed in our case for two main reasons. First, significant spectral broadening requires the gate pulse width to be comparable to the carrier relaxation time in graphene. This condition is not met: our gate pulse width is ~90–100 fs, much shorter than the carrier relaxation time of ~1–2 ps. Second, the laser pulses must propagate through graphene over a sufficiently long distance. In our setup, the gate beam passes only through an ~8-layer graphene flake, whereas in Ref. [56], the pulses propagated along a graphene waveguide, providing a much longer interaction length.

Throughout our optical Kerr gate experiments, multiple flakes and samples, particularly thin graphite films, were investigated using a tunable ultrafast laser system (Taccor Tune 10, Novanta) with a tuning range of 730–880 nm, pulse duration < 80 fs, average power of 2.1 W at 800 nm, and a repetition rate of 1 GHz. The measurements were carried out in a counter-propagating configuration, with the pump (gate) and probe beams being degenerate, namely, operating at the same wavelength. Unlike bulky solid-state Kerr media, identifying a specific number of graphene layers or a thin graphite film of a given thickness requires direct imaging to accurately position the laser spot. To enable this, a white-light source was integrated into the Kerr gate setup to illuminate the sample, and the same 50 mm focal-length lens used to focus the gate beam was also employed for imaging the graphene and graphite films.

Because of the 1 GHz repetition rate, the gate pulse energy is below 1 nJ. To enhance the Kerr gating efficiency despite this low pulse energy, both the probe and gate beams were tightly focused to spot diameters of approximately 20 μm and 30 μm, respectively [25]. The probe beam then passes through a Kerr medium (graphene/graphite film) placed between two crossed polarizers.

In the absence of the gate pulse, the medium remains isotropic (non-birefringent), and the second polarizer blocks the probe beam. When the gate pulse arrives, the optical Kerr effect induces a transient birefringence in the medium, rotating the probe beam's polarization. This rotation allows a portion of the probe light to pass through the second polarizer and be detected by a photodetector. The detected signal corresponds only to the part of the probe pulse that temporally overlaps with the gate pulse. The experimental configuration is shown in Fig. 4.

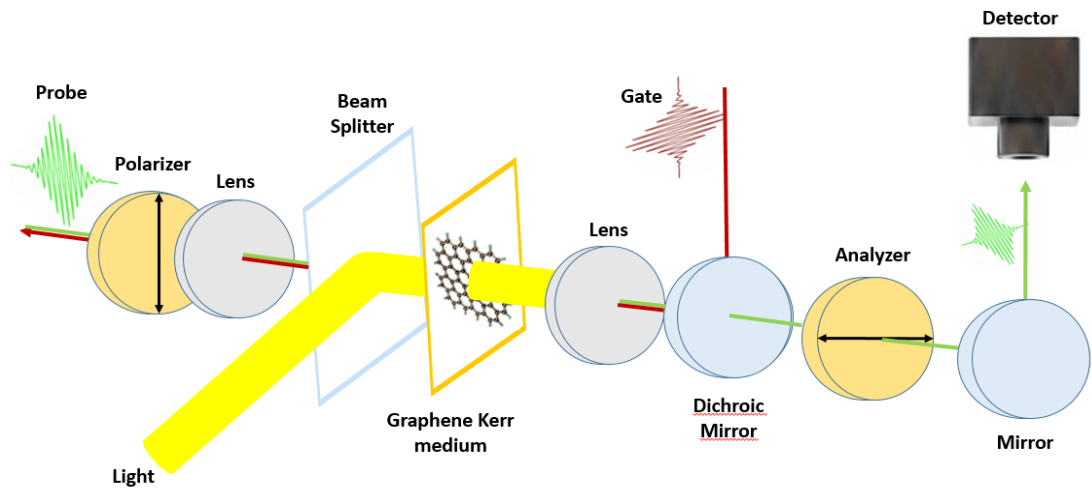

**Figure 4**. Schematics of the experimental setup for OKG experiment.

The gate beam (λ = 820 nm) with an average power ranging from 250–1000 mW (pulse peak intensity ≈ 0.33–1.33 GW/cm²) was scanned across the thin graphite film (~17–18 layers), using a temporal step size of 1 fs. At each delay position, the transmitted Kerr signal was integrated for several seconds. Although a clear time trace was obtained at 1000 mW, the combination of high power, 1 fs step size, and long integration time caused irreversible damage to the sample. By lowering the average power to 250 mW (pulse peak intensity ≈ 0.33 GW/cm²) and increasing the step size to 25 fs, we were able to reliably capture the Kerr response (see Fig. 5 inset). The resulting temporal profile is shown in Fig. 5.

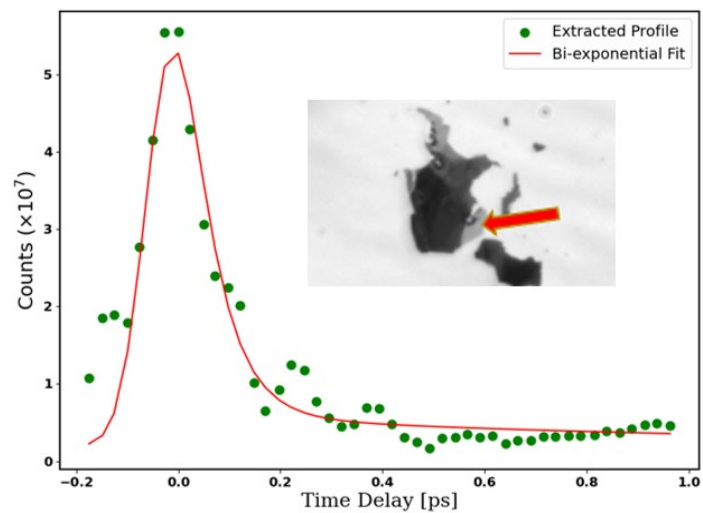

**Figure 5.** Transmitted Kerr signal processed with a Savitzky-Golay filter and fitted using a bi-exponential function convolved with the Gaussian cross-correlation of the pump and probe pulses, yielding a FWHM ≈ 141 ± 6 fs.

The temporal profile shown in Fig. 5 represents the envelope of the OKG signal extracted via Fourier analysis. It reflects the convolution of the cross-correlation between the gate and probe pulses with the intrinsic nonlinear response of the thin graphite film, indicating that the film's nonlinear response time is < 141 fs. A precise determination of the actual response time would require measuring the pump–probe cross-correlation and deconvolving it from the OKG temporal profile. The transmitted

Kerr signal was smoothed using a Savitzky–Golay filter (window of 7 points, polynomial order 3) and fitted with a bi-exponential function convolved with a Gaussian representing the pump–probe cross-correlation, yielding a FWHM ≈ 141 ± 6 fs. The OKG efficiency of our setup was found to be ~ 10 %, evaluated by a separate experiment (not presented here) in which the probe beam was a CW laser at 656 nm, and the gate beam was the fs laser beam at 820 nm, used for obtaining the above results.

## 3. Conclusion

We have developed an ultrafast detection method using the strong third-order optical nonlinearity of graphene and thin graphite films. Operating at a 1 GHz repetition rate with pulse energies below 1 nJ, this platform achieves a temporal resolution of 141 fs, limited primarily by the intrinsic carrier dynamics of the material. The remarkably high nonlinear refractive index of these materials enhances Kerr gating efficiency far beyond conventional media, while their atomic-scale thickness allows straightforward integration with microscopy setups, optical fibers, and on-chip photonic circuits. This compact and versatile approach not only enables sub-picosecond gating and broadband operation but also provides a powerful means to investigate ultrafast photophysical processes at the single-emitter level. Its capabilities make it promising for applications ranging from ultrafast single-photon detection and quantum photonics to high-speed optical switching and time-resolved spectroscopy in both nanophotonic and biological systems.

**Research funding:** This activity has been partially supported by the University of Siegen and the German Research Foundation (DFG) (INST 221/118-1 FUGG, 410405168).

**Author contribution:** All authors have accepted responsibility for the entire content of this manuscript and consented to its submission to the journal, reviewed all the results and approved the final version of the manuscript

**Conflict of interest:** Authors state no conflict of interest.

**Data availability statement**: The datasets generated and analysed during the current study are available from the corresponding author upon reasonable request.

**Acknowledgments:** The authors would like to acknowledge Thomas Lenzer, Kawon Oum, and Giancarlo Soavi for helpful discussions, and Stephan Schäfer for the training on graphene exfoliation and for providing the HOPG mesa.

**References**

1. Y. Arakawa, and M.J. Holmes, "Progress in quantum-dot single photon sources for quantum information technologies: A broad spectrum overview," Applied Physics Reviews 7(2), (2020).

2. V. Shcheslavskiy, P. Morozov, A. Divochiy, Yu. Vakhtomin, K. Smirnov, and W. Becker, "Ultrafast time measurements by time-correlated single photon counting coupled with superconducting single photon detector," Review of Scientific Instruments 87(5), (2016).

3. S. Chan, A. Halimi, F. Zhu, I. Gyongy, R.K. Henderson, R. Bowman, S. McLaughlin, G.S. Buller, and J. Leach, "Long-range depth imaging using a single-photon detector array and non-local data fusion," Sci Rep 9(1), (2019).

4. L. Ma, O. Slattery, and X. Tang, "Single photon frequency up-conversion and its applications," Physics Reports 521(2), 69–94 (2012).

5. E.M.H.P. van Dijk, J. Hernando, J.-J. García-López, M. Crego-Calama, D.N. Reinhoudt, L. Kuipers, M.F. García-Parajó, and N.F. van Hulst, "Single-Molecule Pump-Probe Detection Resolves Ultrafast Pathways in Individual and Coupled Quantum Systems," Phys. Rev. Lett. 94(7), (2005).

6. Y. Venkatesh, M. Venkatesan, B. Ramakrishna, and P.R. Bangal, "Ultrafast Time-Resolved Emission and Absorption Spectra of meso-Pyridyl Porphyrins upon Soret Band Excitation Studied by Fluorescence Up-Conversion and Transient Absorption Spectroscopy," J. Phys. Chem. B 120(35), 9410–9421 (2016).

7. H. Kandori, H. Sasabe, and M. Mimuro, "Direct Determination of a Lifetime of the S2 State of .beta.-Carotene by Femtosecond Time-Resolved Fluorescence Spectroscopy," J. Am. Chem. Soc. 116(6), 2671–2672 (1994).

8. X.-W. Chen, M. Agio, and V. Sandoghdar, "Metallodielectric Hybrid Antennas for Ultrastrong Enhancement of Spontaneous Emission," Phys. Rev. Lett. 108(23), (2012).

9. T.B. Hoang, G.M. Akselrod, and M.H. Mikkelsen, "Ultrafast Room-Temperature Single Photon Emission from Quantum Dots Coupled to Plasmonic Nanocavities," Nano Lett. 16(1), 270–275 (2015).

10. A.M. Flatae, F. Tantussi, G.C. Messina, F. De Angelis, and M. Agio, "Plasmon-Assisted Suppression of Surface Trap States and Enhanced Band-Edge Emission in a Bare CdTe Quantum Dot," J. Phys. Chem. Lett. 10(11), 2874–2878 (2019).

11. S.I. Bogdanov, O.A. Makarova, X. Xu, Z.O. Martin, A.S. Lagutchev, M. Olinde, D. Shah, S.N. Chowdhury, A.R. Gabidullin, I.A. Ryzhikov, I.A. Rodionov, A.V. Kildishev, S.I. Bozhevolnyi, A. Boltasseva, V.M. Shalaev, and J.B. Khurgin, "Ultrafast quantum photonics enabled by coupling plasmonic nanocavities to strongly radiative antennas," Optica 7(5), 463 (2020).

12. B. Schmidt, S. Laimgruber, W. Zinth, and P. Gilch, "A broadband Kerr shutter for femtosecond fluorescence spectroscopy," Appl. Phys. B 76(8), 809–814 (2003).

13. S. Kinoshita, H. Ozawa, Y. Kanematsu, I. Tanaka, N. Sugimoto, and S. Fujiwara, "Efficient optical Kerr shutter for femtosecond time-resolved luminescence spectroscopy," Review of Scientific Instruments 71(9), 3317–3322 (2000).

14. C.H. Kim, and T. Joo, "Ultrafast time-resolved fluorescence by two photon absorption excitation," Opt. Express 16(25), 20742 (2008).

15. O. Kuzucu, F.N.C. Wong, S. Kurimura, and S. Tovstonog, "Time-resolved single-photon detection by femtosecond upconversion," Opt. Lett. 33(19), 2257 (2008).

16. C. Cimpean, V. Groenewegen, V. Kuntermann, A. Sommer, and C. Kryschi, "Ultrafast exciton relaxation dynamics in silicon quantum dots," Laser & Photonics Reviews 3(1–2), 138–145 (2009).

17. L. Zhang, Y.-T. Kao, W. Qiu, L. Wang, and D. Zhong, "Femtosecond Studies of Tryptophan Fluorescence Dynamics in Proteins: Local Solvation and Electronic Quenching," J. Phys. Chem. B 110(37), 18097–18103 (2006).

18. R. Jimenez, G.R. Fleming, P.V. Kumar, and M. Maroncelli, "Femtosecond solvation dynamics of water," Nature 369(6480), 471–473 (1994).

19. A. Barth, "Infrared spectroscopy of proteins," Biochimica et Biophysica Acta (BBA) - Bioenergetics 1767(9), 1073–1101 (2007).

20. M. Glasbeek, and H. Zhang, "Femtosecond Studies of Solvation and Intramolecular Configurational Dynamics of Fluorophores in Liquid Solution," Chem. Rev. 104(4), 1929–1954 (2004).

21. H. Chosrowjan, S. Taniguchi, N. Mataga, M. Unno, S. Yamauchi, N. Hamada, M. Kumauchi, and F. Tokunaga, "Low-Frequency Vibrations and Their Role in Ultrafast Photoisomerization Reaction Dynamics of Photoactive Yellow Protein," J. Phys. Chem. B 108(8), 2686–2698 (2004).

22. J. Xu, and J.R. Knutson, "Chapter 8 Ultrafast Fluorescence Spectroscopy via Upconversion," Methods in Enzymology, 159–183 (2008).

23. W. Tan, H. Liu, J. Si, and X. Hou, "Control of the gated spectra with narrow bandwidth from a supercontinuum using ultrafast optical Kerr gate of bismuth glass," Applied Physics Letters 93(5), (2008).

24. Z. Yu, L. Gundlach, and P. Piotrowiak, "Efficiency and temporal response of crystalline Kerr media in collinear optical Kerr gating," Opt. Lett. 36(15), 2904 (2011).

25. A.-H. Fattah, A.M. Flatae, A. Farrag, and M. Agio, "Ultrafast single-photon detection at high repetition rates based on optical Kerr gates under focusing," Opt. Lett. 46(3), 560 (2021).

26. Y. Huang, E. Sutter, N. N. Shi, J. Zheng, T. Yang, D. Englund, H.-J. Gao, and P. Sutter, Reliable Exfoliation of Large-Area High-Quality Flakes of Graphene and Other Two-Dimensional Materials, ACS Nano 9, 10612 (2015).

27. M. Yi and Z. Shen, A review on mechanical exfoliation for the scalable production of graphene, J. Mater. Chem. A 3, 11700 (2015).

28. P. E. Gaskell, H. S. Skulason, C. Rodenchuk, and T. Szkopek, Counting graphene layers on glass via optical reflection microscopy, Applied Physics Letters 94, (2009).

29. A. C. Ferrari and D. M. Basko, Raman spectroscopy as a versatile tool for studying the properties of graphene, Nature Nanotech 8, 235 (2013).

30. A. C. Ferrari et al., Raman Spectrum of Graphene and Graphene Layers, Phys. Rev. Lett. 97, (2006).

31. J. Yan, Y. Zhang, P. Kim, and A. Pinczuk, Electric Field Effect Tuning of Electron-Phonon Coupling in Graphene, Phys. Rev. Lett. 98, (2007).

32. A. C. Ferrari, S. E. Rodil, and J. Robertson, Interpretation of infrared and Raman spectra of amorphous carbon nitrides, Phys. Rev. B 67, (2003).

33. A. C. Ferrari, Raman spectroscopy of graphene and graphite: Disorder, electron–phonon coupling, doping and nonadiabatic effects, Solid State Communications 143, 47 (2007).

34. C. H. Lui, Z. Li, Z. Chen, P. V. Klimov, L. E. Brus, and T. F. Heinz, Imaging Stacking Order in Few-Layer Graphene, Nano Lett. 11, 164 (2010).

35. Y. Yang et al., Stacking Order in Graphite Films Controlled by van der Waals Technology, Nano Lett. 19, 8526 (2019).

36. C. Cong, T. Yu, K. Sato, J. Shang, R. Saito, G. F. Dresselhaus, and M. S. Dresselhaus, Raman Characterization of ABA- and ABC-Stacked Trilayer Graphene, ACS Nano 5, 8760 (2011).

37. K. G. Wirth et al., Experimental Observation of ABCB Stacked Tetralayer Graphene, ACS Nano 16, 16617 (2022).

38. S. Katsiaounis, N. Chourdakis, E. Michail, M. Fakis, I. Polyzos, J. Parthenios, and K. Papagelis, Graphene nano-sieves by femtosecond laser irradiation, Nanotechnology 34, 105302 (2022).

39. C. Berger et al., Electronic Confinement and Coherence in Patterned Epitaxial Graphene, Science 312, 1191 (2006).

40. X. Fan, G. Zhang, and F. Zhang, Multiple roles of graphene in heterogeneous catalysis, Chem. Soc. Rev. 44, 3023 (2015).

41. W. Yuan and G. Shi, Graphene-based gas sensors, J. Mater. Chem. A 1, 10078 (2013).

42. H. O. Jeschke, M. E. Garcia, and K. H. Bennemann, Theory for the Ultrafast Ablation of Graphite Films, Phys. Rev. Lett. 87, (2001).

43. A. Beltaos, A. Kovačević, A. Matković, U. Ralević, D. Jovanović, and B. Jelenković, Damage effects on multi-layer graphene from femtosecond laser interaction, Phys. Scr. T162, 014015 (2014).

44. A. Roberts, D. Cormode, C. Reynolds, T. Newhouse-Illige, B. J. LeRoy, and A. S. Sandhu, Response of graphene to femtosecond high-intensity laser irradiation, Applied Physics Letters 99, (2011).

45. M. Currie, J. D. Caldwell, F. J. Bezares, J. Robinson, T. Anderson, H. Chun, and M. Tadjer, Quantifying pulsed laser induced damage to graphene, Applied Physics Letters 99, (2011).

46. B. Krauss, T. Lohmann, D.-H. Chae, M. Haluska, K. von Klitzing, and J. H. Smet, Laser-induced disassembly of a graphene single crystal into a nanocrystalline network, Phys. Rev. B 79, (2009).

47. F. Schedin, A. K. Geim, S. V. Morozov, E. W. Hill, P. Blake, M. I. Katsnelson, and K. S. Novoselov, Detection of individual gas molecules adsorbed on graphene, Nature Mater 6, 652 (2007).

48. M. Kaiser, M. Kahmann, J. Kleiner, and D. Flamm, Tailored-edge laser glass cleaving supported by thermal separation, Laser Applications in Microelectronic and Optoelectronic Manufacturing (LAMOM) XXVIII 26 (2023).

49. M. Kaiser, J. Kleiner, J. Wolff, and D. Flamm, Customized edge cutting of display glass with laser-only machining, Frontiers in Ultrafast Optics: Biomedical, Scientific, and Industrial Applications XXIV 27 (2024).

50. S. A. Mikhailov, Non-linear electromagnetic response of graphene, Europhys. Lett. 79, 27002 (2007).

51. S. A. Mikhailov and K. Ziegler, Nonlinear electromagnetic response of graphene: Frequency multiplication and the self-consistent-field effects, J. Phys.: Condens. Matter 20, 384204 (2008).

52. Q. Bao, H. Zhang, Y. Wang, Z. Ni, Y. Yan, Z. X. Shen, K. P. Loh, and D. Y. Tang, Atomic-Layer Graphene as a Saturable Absorber for Ultrafast Pulsed Lasers, Adv Funct Materials 19, 3077 (2009).

53. E. Hendry, P. J. Hale, J. Moger, A. K. Savchenko, and S. A. Mikhailov, Coherent Nonlinear Optical Response of Graphene, Phys. Rev. Lett. 105, (2010).

54. J. L. Cheng, N. Vermeulen, and J. E. Sipe, Third order optical nonlinearity of graphene, New J. Phys. 16, 053014 (2014).

55. D. Castelló-Lurbe, H. Thienpont, N. Vermeulen, Predicting Graphene's Nonlinear-Optical Refractive Response for Propagating Pulses. Laser & Photonics Reviews, 14, 1900402 (2020).

56. K. L. Ishikawa, Nonlinear optical response of graphene in time domain, Phys. Rev. B 82, (2010).55. Vermeulen, N., Castelló-Lurbe, D., Khoder, M. et al. Graphene's nonlinear-optical physics revealed through exponentially growing self-phase modulation. Nat Commun 9, 2675 (2018).